\documentclass[conference,]{IEEEtran}

\usepackage[utf8]{inputenc}
\usepackage{sistyle}
\usepackage[english]{babel}
\usepackage[english]{isodate}
\usepackage{cmap}
\usepackage[T1]{fontenc}

\usepackage{cite}
\usepackage{graphicx}
\usepackage[cmex10]{amsmath}

\usepackage{array}

\usepackage{mdwmath}
\usepackage{mdwtab}

\usepackage{algpseudocode}

\usepackage[bookmarks=false]{hyperref}
\hypersetup{pdfhighlight=/N,pdfborder={0 0 0}}
\usepackage[toc=false,acronym]{glossaries}
\newacronym{pbc}{PBC}{periodic boundary conditions}
\newacronym{asep}{ASEP}{asymmetric exclusion process}
\newacronym{sca}{SCA}{stochastic cellular automaton}
\newacronym{mcs}{MCS}{Monte-Carlo step}
\newacronym{nn}{NN}{nearest neighbor}
\newacronym{simd}{SIMD}{single instruction multiple data}
\newacronym{simt}{SIMT}{single instruction multiple thread}
\newacronym{lsb}{LSB}{least significant bit}
\newacronym{gpu}{GPU}{graphics processing unit}
\newacronym{lcg}{LCG}{linear congruential generator}
\newacronym{tdp}{TDP}{thermal design power}
\newacronym{kpz}{KPZ}{Kardar--Parisi--Zhang}

\makeglossaries

\usepackage{calc}

\newcommand{\slope}[2]{\ensuremath{\sigma_{#1^{#2}}}}
\newcommand{\vslope}[2]{\vec\sigma_{#1^{#2}}}

\usepackage{tikz}
\usepackage{textcomp}
\usepackage{hyperref}
\usepackage{lipsum}

\newcommand\copyrighttext{%
  \footnotesize \textcopyright 2016 IEEE. Personal use of this material is permitted.
  Permission from IEEE must be obtained for all other uses, in any current or future
  media, including reprinting/republishing this material for advertising or promotional
  purposes, creating new collective works, for resale or redistribution to servers or
  lists, or reuse of any copyrighted component of this work in other works.
  DOI: \href{}{}}
\newcommand\copyrightnotice{%
\begin{tikzpicture}[remember picture,overlay]
\node[anchor=south,yshift=10pt] at (current page.south) {\fbox{\parbox{\dimexpr\textwidth-\fboxsep-\fboxrule\relax}{\copyrighttext}}};
\end{tikzpicture}%
}

\begin{document}
\glsdisablehyper

\title{Bit-Vectorized \glsname{gpu} Implementation of\\
a Stochastic Cellular Automaton Model\\
for Surface Growth}


\author{\IEEEauthorblockN{%
 Jeffrey Kelling\IEEEauthorrefmark{1},
 {G\'eza \'Odor}\IEEEauthorrefmark{2}, 
 {Sibylle Gemming}\IEEEauthorrefmark{3}\IEEEauthorrefmark{4}
}
\IEEEauthorblockA{\IEEEauthorrefmark{1}%
Helmholtz-Zentrum Dresden - Rossendorf,
Department of Information Services and Computing, \\
Bautzner Landstra\ss{}e 400, 01328 Dresden, Germany \\
Email: j.kelling@hzdr.de}
\IEEEauthorblockA{\IEEEauthorrefmark{2}%
MTA-EK-MFA Institute of Technical Physics and Materials Science,
Budapest, Hungary
}
\IEEEauthorblockA{\IEEEauthorrefmark{3}%
 Helmholtz-Zentrum Dresden - Rossendorf,
Institute of Ion Beam Physics and Materials Research, \\
Bautzner Landstra\ss{}e 400, 01328 Dresden, Germany
}
\IEEEauthorblockA{\IEEEauthorrefmark{4}%
Institute of Physics, TU Chemnitz,
09107 Chemnitz, Germany}
}


%


\maketitle
\copyrightnotice

\begin{abstract}
Stochastic surface growth models aid in studying properties of universality
classes like the Kardar--Parisi--Zhang class. High precision results obtained
from large scale computational studies can be transferred to many physical
systems. Many properties, such as roughening and some two-time functions can be
studied using \gls{sca} variants of stochastic models. Here we present a highly
efficient \gls{sca} implementation of a surface growth model capable of
simulating billions of lattice sites on a single GPU. We also provide insight
into cases requiring arbitrary random probabilities which are not accessible
through bit-vectorization.
\end{abstract}

%
\IEEEpeerreviewmaketitle

\glsreset{sca}

\section{Introduction}

The roughening and aging of interfaces is important in many processes in physics
and technology. Examples are the progression of solidification
fronts~\cite{langer1980} or flame
fronts~\cite{landau1984book} and the growth of surfaces.  The
\gls{kpz} equation~\cite{PhysRevLett.56.889} is a stochastic
differential equation devised to model kinetic roughening of surfaces under
ballistic deposition of particles.  It describes the evolution of the
surface height as an interplay of linear growth and processes of diffusion,
smoothing the surface, and a local instability, increasing local roughness,
with random particle deposition noise:
\begin{equation}
 \partial_t h(\mathbf x, t) =  \underbrace{v}_\text{lin.}
 + \underbrace{\nu\nabla^2 h(\mathbf x, t)}_\text{diffusion}
 + \underbrace{\lambda[\nabla h (\mathbf x, t)]^2}_\text{roughening}
 + \underbrace{\eta(\mathbf x, t)}_\text{noise}
 \label{eq:kpz}
\end{equation}
Here, $t$ is the time and $\mathbf x$ the position on the surface. $\eta$
denotes an uncorrelated Gaussian noise with zero mean.

This equation was also found to model different systems like randomly stirred
fluids~\cite{forster77}, directed polymers in random media~\cite{Kardar87} and
even magnetic flux lines in superconductors~\cite{hwa92}. The concept of aspects
of different physical systems exhibiting the same scaling behavior is called
universality~\cite{odorbook}. Solutions of the \gls{kpz} equation define
universality classes.

Exact solutions are available only in one-dimension.  The growth of
two-dimensional surfaces into the third dimension ($2+1$ dimensions) is of
special interest due to the experimental relevance. Thus experimental
studies are performed to compare empirical surface growth with \gls{kpz}
universality.~\cite{halpinhealy2014}
On the other hand computational models which fall into this universality class
are used to gather insight into general properties of these physical systems.

In order to compare calculations to experiments, large scale simulations, both
in time and space are advantageous. A variety of stochastic surface growth
models has been devised to accomplish this task. For example the restricted
solid-on-solid model~\cite{kimKosterlitz1989} or the $d$-mer
model~\cite{odor09,PhysRevE.81.031112}, which is the subject of the present work.

Efficient implementation of stochastic lattice models has been of interest
for a very long time, primarily for spin models like the Ising
model~\cite{landau05a}. The most efficient way is to formulate the model as
a \gls{sca}~\cite{wolfram2002new}. Ever since \glspl{gpu} have been introduced
they have also been employed for these tasks~\cite{preis09,KONSH2012}. A very
recent study of bit-vectorized Ising spin-glass simulations on \gls{gpu} can be
found in~\cite{LBP2015}. 

The one-dimensional $d$-mer model (roof-top model)~\cite{PhysRevB.35.3485} has
also been studied
using a \gls{sca} of an \gls{asep}~\cite{SOON2011}, but for limited system sizes.
Here we present a very efficient bit-vectorized \gls{sca} implementation of the
$2+1$ dimensional $d$-mer model, which scales to very large
systems ($\sim 10^{10}$ lattice sites on a single \gls{gpu}). Dealing with
arbitrary update probabilities is also discussed here, which is often not
considered in bit-vectorized implementations.

In the following, the $2+1$ dimensional $d$-mer model, the octahedron model, is
introduced.
Section~\ref{s:impl} describes our bit vectorized \gls{sca} implementation, the
performance of which is analyzed in section~\ref{s:perf}. Finally, conclusions are
presented in section~\ref{s:concl}.

\subsection{The Octahedron Model on a Checkerboard}

\begin{figure}
 \centering
 \includegraphics[width=\linewidth*2/3]{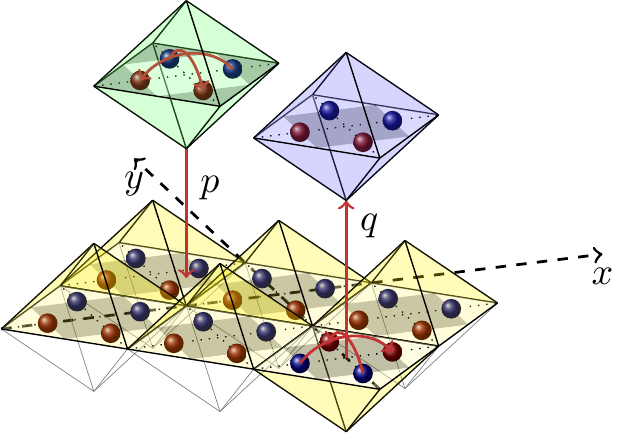}
 \caption{Schematic of allowed update processes in the octahedron model. Dark
  and light squares mark even and odd lattice sites, respectively. Up and down
  slopes between sites are represented by red and blue balls, respectively.
  Deposition or removal processes, carried out with probabilities $p$ or
  $q$ respectively, involve exchanging
  pairs of up and down slopes along $x$ and $y$ direction.
  }
 \label{fig:octahedronModel}
\end{figure}

A growing surface can be most efficiently represented by subtracting the
linearly growing average height and considering only local height differences,
which govern the surface morphology. In the octahedron model these height
differences (slopes) are restricted to $\slope{x/y}{} =\pm 1$. Updates,
corresponding to the deposition or removal of particles on the surface, follow
the generalized Kawasaki rules:~\cite{odor09,PhysRevE.81.031112}
\begin{equation}
\left(
\begin{array}{cc}
   -1 & 1 \\
   -1 & 1 
\end{array}
\right)
 \overset{p}{\underset{q}{\rightleftharpoons }}
\left(
\begin{array}{cc}
   1 & -1 \\
   1 & -1 
\end{array}
\right) \label{eq:kawasakiRules}
\quad,
\end{equation}
where allowed deposition processes are carried out with probability $p$ and
removals with probability $q$. These rules are an extension of an \gls{asep} in
one dimension.~\cite{PhysRevB.35.3485} Figure~\ref{fig:octahedronModel}
illustrates the update processes as exchanges of slopes along the $x$ and $y$
axes. \Gls{pbc} are applied, that is slopes leaving the system at one edge
reenter at the opposite edge.

A time step, or \gls{mcs}, requires attempting one update for each lattice site,
this constitutes the unit of time in the simulation.  To achieve this in a
well-defined manner, the \gls{sca} first updates the even (black) lattice sites,
then the odd (white) ones as defined by the parity of the xor of the coordinates
$(x \oplus y) \land 1$.

The focus of the presented implementation lies on cases where $p>0, q=0$. Using
these parameters, the surface growth will follow \gls{kpz}
universality exhibiting power-law growth of the surface roughness with a
non-trivial exponent.~\cite{PhysRevE.84.061150} In these cases, the choice of
$p$ rescales the simulation time non-linearly and affects correlation functions,
beyond a simple rescaling of time. If $p=q$ the resulting surface growth falls
in the Edwards--Wilkinson universality class, where the roughness grows
logarithmically~\cite{EW,odor09}.

\section{Implementation\label{s:impl}}

\subsection{Lattice Encoding}

\begin{figure}
 \centering
 \includegraphics[width=\linewidth]{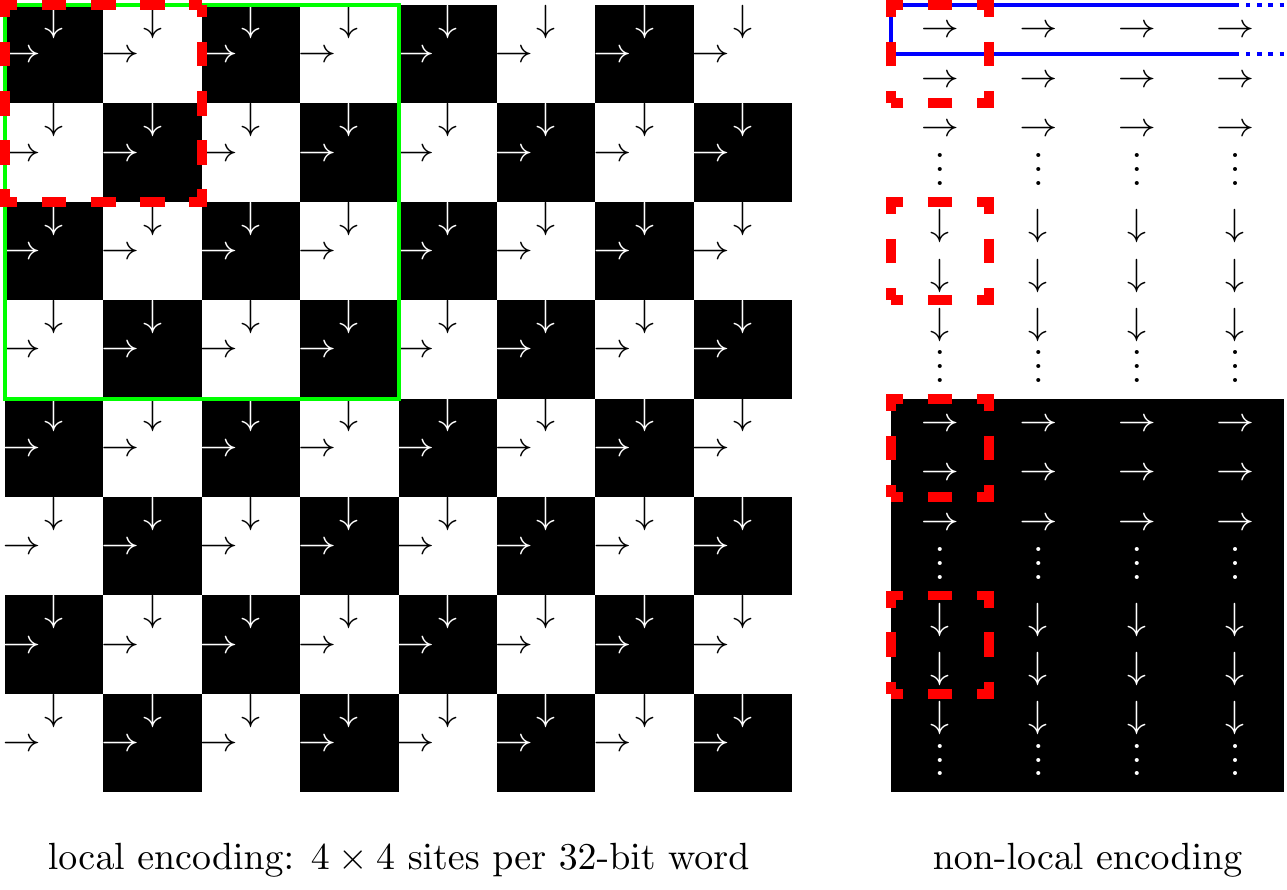}
 \caption{Comparison of direct, local bit-coding (left) and non-local encoding
  suitable for vectorization. Black and white areas represent even and odd
  lattice sites, respectively. Arrows represent slopes connecting neighbors in
  the indicated direction. Red dashed frames show the correspondence between
  locally and non-locally encoded slope information. Solid frames indicate data
 stored in a 32-bit word.}
 \label{fig:octNlEncoding}
\end{figure}

The slopes can be mapped to a binary state $\slope{x/y}{} \to
\{0,1\}$ of which two need to be encoded per lattice site ($\slope x-$ and
$\slope y-$). The slopes to the remaining neighbors can be retrieved from those:
$\slope x+(x,y) = \slope x-(x+1,y)$ and $\slope y+(x,y) = \slope y-(x,y+1)$ to avoid
redundancy. Storing two bits per lattice site and encoding tiles of $4\times4$
in 32-bit words is straightforward. This encoding is illustrated in the left-hand panel
of figure~\ref{fig:octNlEncoding}. It has the advantage that \gls{nn} sites,
which are required to perform an update may be found in the same word, making it
the encoding of choice if lattice sites are selected
randomly~\cite{PhysRevE.84.061150} and vectorization is not an option.

For bit-vectorization, the ideal encoding is non-local with respect to the
relations of data in configuration space. Instead the bits are grouped in memory
according to their role. The \gls{sca} defines four roles in total, based on two
binary properties: The direction of the slope $\slope{x/y}{}$ and the parity of
the lattice site (odd/even). This grouping is illustrated in the right-hand
panel of figure~\ref{fig:octNlEncoding}: All slopes with the same role are
stored at consecutive addresses in memory, allowing vectorization with any word
size not larger than half the size of the simulation cell $X/2$.

\subsection{Bit-Vectorized \gls{gpu} Implementation}

The kernel of the bit-vectorized \gls{gpu} implementation is summarized in pseudo-code
in figure~\ref{fig:octNlAlgo}. Care must be taken regarding the parity of rows
with respect to the considered sub-lattice (line~\algref{fig:octNlAlgo}{alg:parRow}): All
slopes belonging to sites in odd rows are perfectly aligned and no shifting is
required. In even rows the $\slope x+$ slopes, stored at the \gls{nn} site in
$x$ direction, are shifted by one bit in memory (compare
figure~\ref{fig:octNlEncoding}, left).

The implementation treats consecutive words processed by threads of the same
block as one effective \gls{simd} word, with a maximum effective size
of $w_\mathrm{eff,max} = w\times{}maxThreadsPerBlock$, where $w$ is the word size
in bits, which is 32 on \gls{gpu}, giving $w_\mathrm{eff,max}=2^{15}$. This
\gls{simd}-word size can be adjusted to span the simulation cell, as long as the
lateral size does not exceed $X=2^{16}$. Larger simulations are rarely required,
thus the kernel displayed in figure~\ref{fig:octNlAlgo} assumes the
effective \gls{simd} word to span the system. The work assigned to thread blocks
is distributed along the $y$ direction.  Buffer regions between blocks or global
atomics are not employed since, as long as all blocks update the same
sublattice, the non-locality of the encoding of on-site slopes avoids write
conflicts in global memory.

\begin{figure}
 \centering
 \begin{algorithmic}[1]
  \Require $X\times Y$ \Comment{system size}
  \Require $w$ \Comment{word size in bits}
  \Require $\vec\sigma$ \Comment{bit-vector of slopes, one word}
  \Require $\vslope{x/y}-^{0/1}[X/w/2, Y]$ \Comment{4 arrays of
  slopes ($x/y$, even/odd)}
  \Statex
   \For{$y\in blockBounds$}
    \State $par_\mathrm{row}\gets par_\mathrm{lat} \otimes
    \Call{parity}{y}$\label{alg:parRow}
    \State $x_w \gets t_{id}$\label{alg:xw}
     \Comment{index of slope-vector in $x$}
    \Statex
     \State $\vslope{x/y}- \gets \vslope {x/y}-^{par_\mathrm{lat}}[x_w, y]$
     \State $\vslope y+ \gets \vslope y-^{\lnot par_\mathrm{lat}}[x_w, y+1]$
      \Comment{\gls{nn} $y$}
     \State $\vslope x+ \gets \vslope x-^{\lnot par_\mathrm{lat}}[x_w, y]$
      \Comment{\gls{nn} $x$}
     \State $\vslope x+^\mathrm{shrd}[t_{id}] \gets \vslope x+$
     \Comment{shared buffer for rotation}
     \Statex
     \If{$\lnot par_\mathrm{row}$}\label{alg:ifEvenRowForth}
      \Comment{rotate \glsname{simd} word by one bit:}
      \State \Call{synchronizeThreads}{}
      \State $\vslope x+ \gets\Call{shiftRight}{\vslope x+,1}$
      \State $\vslope x+ \gets{\vslope x+} \lor \Call{shiftLeft}{\vslope
      x+^\mathrm{shrd}[t_{id} + 1], w-1}$
     \EndIf
     \Statex\\
     \Comment{compute xor mask for update:}
     \State $m \gets \Call{updateMask}{\vslope x-,\vslope y-,\vslope
     x+,\vslope y+, p, q}$ \label{alg:updateMask}
     \Statex\\
     \Comment{apply mask:}
     \State $\vslope {x/y}-^{par_\mathrm{lat}}[x_w, y]    \gets m \oplus \vslope {x/y}-^{par_\mathrm{lat}}[x_w, y]$
     \State $\vslope y-^{\lnot par_\mathrm{lat}}[x_w, y+1]\gets m \oplus \vslope y-^{\lnot par_\mathrm{lat}}[x_w, y+1]$
     \Statex
     \If{$\lnot par_\mathrm{row}$}\label{alg:ifEvenRowBack}\\
      \Comment{apply mask to \gls{nn} $x$ slopes, except \glsname{lsb}:}
      \State $\vslope x+^\mathrm{shrd}[t_{id}] \gets \vslope
      x+^\mathrm{shrd}[t_{id}] \oplus \Call{shiftLeft}{m,1}$
      \State \Call{synchronizeThreads}{}\label{alg:ifEvenRowBackSync1}\\
      \Comment{apply mask to \glsname{lsb} of next \gls{nn} $x$ slopes:}
      \State $\vslope x+^\mathrm{shrd}[t_{id}+1] \gets \vslope
      x+^\mathrm{shrd}[t_{id}] \oplus \Call{shftRight}{m,w-1}$
      \State \Call{synchronizeThreads}{}
      \State $\vslope x+^{\lnot par_\mathrm{lat}}[x_w, y]  \gets \vslope x+^\mathrm{shrd}[t_{id}]$
     \Else
      \State $\vslope x+^{\lnot par_\mathrm{lat}}[x_w, y]  \gets m \oplus
      \vslope x+$
     \EndIf
     \Statex
   \EndFor
 \end{algorithmic}
 \caption{Bit-vectorized \gls{gpu} kernel using non-local encoding to update one
  sublattice with given parity ($par_\mathrm{lat} = (x \oplus y) \land 1$). It
  is executed for each sublattice in order to complete one \gls{mcs}.  \gls{pbc}
  apply for all coordinates, including indexes in the shared memory buffer
  $\vslope x+^\mathrm{shrd}[]$. $t_\mathrm{id}$ is a short-hand for the
  thread~ID within the thread block. The above directly applies if the thread block
  ($\hat=$ \gls{simd} word) spans the system in $x$ direction.
 Statements involving $\vslope{x/y}-$ represent two separate operations on
$\vslope x-$ and $\vslope y-$. Load and store operations for parameters and
states of random number generators happen before respectively after the
presented loop and are omitted for brevity. See text for details.}
 \label{fig:octNlAlgo}
\end{figure}

Each thread updates 32 lattice sites simultaneously. The corresponding slopes
$\vslope x+$ are placed in a buffer in shared memory ($\vslope
x+^\mathrm{shared}$) to facilitate block-wide rotation when updating even
rows. The correct set of $\vslope x+$ slopes is compiled in the lines
following~\algref{fig:octNlAlgo}{alg:ifEvenRowForth}. An xor-mask $m$ encoding
the Kawasaki exchanges to be carried out is used to apply updates in parallel.
Again, $\vslope x+$ needs to be treated separately in even rows: The update mask
is applied in parts to the corresponding data in shared memory. Atomics could be
used to apply $m$, in order to remove the need for synchronization in
line~\algref{fig:octNlAlgo}{alg:ifEvenRowBackSync1}, but this was found to
yield lower performance.

The update mask $m$ in line~\algref{fig:octNlAlgo}{alg:updateMask} is generated
according to the Kawasaki rules in equation~\eqref{eq:kawasakiRules}, which can
be implemented by the following relations:
\begin{align}
 m_p  &= \vec\xi_p \land \lnot(\vslope x- \lor \vslope y-) \land \vslope x+ \land
 \vslope y+ \\
 m_q  &= \vec\xi_q \land \lnot(\vslope x+ \lor \vslope y+) \land \vslope x- \land
 \vslope y- \\
 m &= m_p \oplus m_q \quad,
\end{align}
where $\vec\xi_r$ denotes a word of random bits set with probability~$r$. If
$q=0$, the calculation of $m_q$ can be omitted and $m=m_p$.

Generation of $\vec\xi_{0.5}$ is trivial and fastest, provided a good pseudo-random
number generator with all bits with uniform distribution. For an arbitrary $r$, it
is necessary to generate $\vec\xi_r$ sequentially using $w$ random numbers. All
$w=32$ random bits need to be generated.  Generating only those random
bits which are required for the actually possible updates in a way which minimizes
warp-divergence, does not improve performance. Even then the implementation is
considerably faster than a non-vectorized version using local encoding.

With the effective \gls{simd} word spanning the simulation cell in one
direction, the implementation utilizes global memory perfectly efficiently: All
accesses are coalesced and all bits read are required to perform updates. To
handle even larger systems ($X\geq2^{17}$), the kernel needs to be changed. The
word index $x_w$, initialized in line~\algref{fig:octNlAlgo}{alg:xw}, is
iterated with a stride equaling the number of threads per block. The procedure
remains unchanged for odd rows, where no slopes need to be shifted. For even
rows, the shared buffer $\vslope x+^\mathrm{shared}$ holds two \gls{simd} words.
$\vslope x+$ slopes need to be shifted between neighboring \gls{simd} words,
thus the whole of a second \gls{simd} word is cached to ensure keeping global
memory accesses coalesced. The buffer is used in a wrap-around fashion ($\vslope
x+^\mathrm{shared}[x_w \mod bufferSize ]$) and the first word $\vslope x+[0]$ is
cached in shared memory separately in order to apply \glspl{pbc} without having to
read the data from global memory twice.

Good quality pseudo-random numbers are crucial in this implementation,
especially in the optimized case $r=0.5$. While in some cases adequate for
Monte-Carlo methods, a \gls{lcg} is insufficient here: Correlations would
severely influence results and produce accumulating errors since the \gls{sca}
updating procedure is itself correlated and is to be decorrelated by the random
acceptance of updates in the first place. Also, \glspl{lcg} do usually not
provide good randomness of single bits. The present implementation uses a small
version of the well-known Mersenne Twister~\cite{matsumoto1998mersenne}, called
TinyMT~\cite{tmt}, which was developed by the same authors. Each thread
maintains its own, randomly seeded, TinyMT state of 128 bits and a polynomial of
96 bits. The polynomials of all threads are independent of each other.

For performance comparisons, a multi-threaded CPU implementation using a word
size $w=64$ bit was created for the case $p=0.5, q=0$. The usage of \gls{simd}
instructions through the vector extensions of GCC~\cite{gccvecex} did not
increase performance. Probably because some operations, like bit shifts across
a whole \gls{simd} word, require more operations than with scalar commands,
compensating the performance gain from vectorization.

\section{Performance\label{s:perf}}

\begin{table}
 \centering
 \renewcommand{\arraystretch}{1.2}
 \caption{\label{tab:hardware}
  List of compute devices used for benchmarks.
 }
 \begin{tabular}{lrrr}
  \hline
  Device & Cores & Max. Bandwidth [GB/s] & TDP [W]\\
  \hline
  Intel i7-4930K & 6(12) & 59.7 & 130\\ 
  GTX Titan Black & 2880 & 336 & 250 \\
  K80 (1 \gls{gpu}) & 2496 & 240 & 150\\
  GTX Titan X & 3072 & 336.5 & 250\\
  \hline
 \end{tabular}
\end{table}

The devices used for performance measurements are listed in
table~\ref{tab:hardware}. Tests of the CPU implementation were performed by
running the maximum number of hardware threads provided by the hyper-threading
capabilities of the platform (12).
This increases the performance by less than one percent over running only one
thread per physical core (6) for the bit-vectorized implementation. This suggests
that the performance of the bit-vectorized implementation on CPU is not
significantly limited by memory latency. The gain is about 20\% for the local
implementation. 

\begin{figure}
 \centering
 \includegraphics[width=\linewidth]{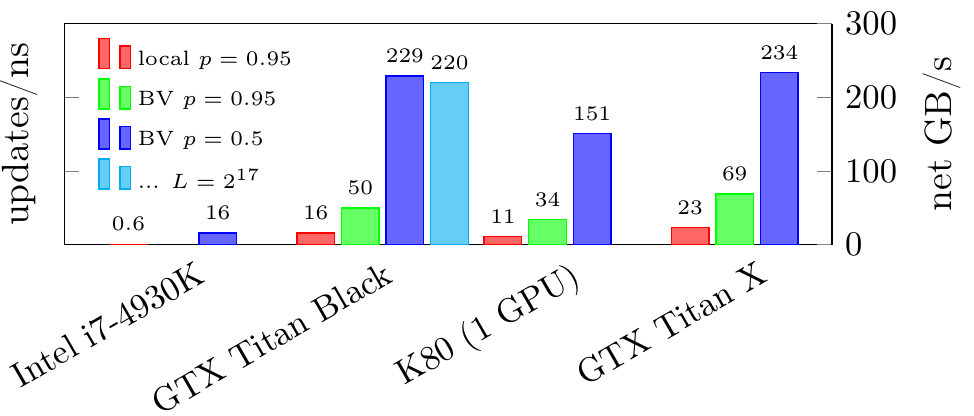}
 \caption[\gls{sca} octahedron model benchmarks]{\label{fig:parallelSCABench}%
 Benchmark results of various \gls{sca} implementations of the octahedron model
 on Kepler (GTX TitanBlack, K80) and Maxwell (GTX TitanX) generation NVIDIA
 \glspl{gpu} and on an Intel i7-4930K CPU. The benchmarks on \glspl{gpu} were
 performed for systems of lateral size $L=X=Y=2^{16}$ (except for the cyan bar,
 where $L=2^{17}$, see text). Benchmarks on CPU used a lateral size of $L=2^{14}$
 lattice sites. All presented benchmarks were performed for $q=0$, $p$ as given
 in the legend. The right axis translates the rate of performed updates into the
 net bandwidth required to read and write the processed data, excluding any
 overhead.
}%
\end{figure}

The achieved performance of different implementations is presented in
figure~\ref{fig:parallelSCABench} as the number of update attempts performed per
nanosecond. In the course of one \gls{mcs}, each slope needs to be touched
twice, translating into two write and two read accesses to each slope per \gls{mcs}
in the ideal case. This is illustrated by the right axis of the plot associating
the performance with the ideally required bandwidth. For $p=0.5,q=0$, the
bit-vectorized implementation performs about 229 updates per nanosecond on a GTX
Titan Black \gls{gpu}, which requires 229 GB/s of slope data to be transferred
between global device memory and the \gls{gpu}. The NVIDIA Profiler
\texttt{nvprof} reports an achieved memory bandwidth of about the same value,
since the memory transfers required to load and store states and parameters of
random number generators negligible. This bandwidth is lower than the maximum
bandwidth listed for the device in table~\ref{tab:hardware}, but it is equal to
the bandwidth reported by the \texttt{bandwithTest} utility from the CUDA SDK
for pure device-to-device memory copy. At the same time \texttt{nvprof} reports
low to medium utilization of arithmetic units. Thus the code is clearly
memory-bound.

When the lateral system size is increased to $X=2^{17}$, the effective
\gls{simd} word does no longer span the simulation cell and memory accesses
cannot be done as efficiently anymore, this leads to a performance drop of
less than 5\% (cyan bar).

In the case of arbitrary $p$, the generation of correctly distributed random
bits becomes more expensive. The benchmarks for $p=0.95$ (green bars in
figure~\ref{fig:parallelSCABench}) show a performance reduced by a factor of
four to
five, which is still faster than the local implementation for the same case (red
bars). Since memory access patterns of the bit-vectorized implementation remain
unchanged between the cases $p=0.50$ and arbitrary $p$, the decrease in
performance can only stem from the increased computational load of random number
generation, which does not require branches. This means, that the implementation
turns from memory-bandwidth-bound to compute bound.  Specific choices for the
probabilities, which
are the sums or differences of fractions $1/2^i$ can be generated at lower
computational cost than arbitrary ones, since they can be generated from
less than 32 random numbers using the logical ``and'' and ``or'' operations.

The newer GTX Titan X (Maxwell-generation)
\gls{gpu} provides a significant speedup over Titan Black for arbitrary $p$, due
to increased compute power, but only marginal gain in the case $p=0.5$, since
the available memory bandwidth is almost the same.

The values listed for \gls{tdp} in table~\ref{tab:hardware} may
not be perfectly comparable across vendors due to different definitions, but can
provide a rough estimate of the power consumption of the device in a steady
state under load. The actual power consumption during the benchmarking was not
measured. Comparing performance and \gls{tdp} for the Kepler-generation cards,
one can conclude that the lower clocked, compute \gls{gpu} K80 is more energy
efficient running the presented implementations than the gaming card.

The case $p=q=0.5$ was also tested, but not separately optimized, because
simulations of this case are of less scientific interest since it falls into the
analytically-solved Edwards--Wilkinson universality class. However, evaluating
additional $q=0.5$ updates does not significantly affect performance since
memory bandwidth remains the main limiting factor.

\section{Conclusion\label{s:concl}}

A very efficient \gls{sca} implementation of the octahedron model describing
surface
growth was presented, which is capable of simulating large systems over long
simulation times (for example $L=2^{16}$ at $p=0.5$ for 1.4M\gls{mcs} in under
twelve hours using a K80 \gls{gpu}). This implementation enables large scale
studies of the physical aging behavior of the surface growth models under
\gls{sca} dynamics~\cite{kpzAging2016}.

It was shown that the code is memory-bound on \glspl{gpu} in the special case
of $p=0.5$. When using an arbitrary update probability, where a separate random
number is required for each lattice site, the implementations performance drops
only by a factor of four to five. Thus it was shown, that by using bit-vectorized
updates and an optimized encoding scheme, the performance of the \gls{sca} can
be increased even without restricting possible applications by the
site-updating frequencies.  In this case the performance is limited by the
performance of random number generation, which might be optimized, depending on
the requirements of calculations.

A fair comparison was done between the performance on CPUs and \glspl{gpu}, since
implementations on both architectures are highly optimized. A speedup of up to
$14\times$ was found for \gls{gpu} over a single socket six core CPU.

The program presented here will be made available upon request to the
corresponding author.

\section*{Acknowledgment}

We thank the HZDR computing center and M. Bussmann for providing computational
resources. We acknowledge the Center for Information Services and High
Performance Computing (ZIH) at TU Dresden for providing us with access to the
Taurus cluster and NIIF for awarding us access to clusters in Budapest and
Debrecen (Hungary),
This work has been partially funded by the Initiative and Networking Fund of the
German Helmholtz Association via the W2/W3 Programm f\"ur exzellente
Wissenschaftlerinnen \mbox{(W2/W3-026)} and the Helmholtz International Research
School for Nanoelectronic Networks NanoNet \mbox{(VH-KO-606)}. We gratefully
acknowledge further funding by the GCoE Dresden. G.~\'O.~acknowledges \mbox{OKTA
(K10 9577)} fund support.



%
\newpage
 \bibliography{IEEEtranBST2/IEEEabrv,bib}
 \bibliographystyle{IEEEtranBST2/IEEEtran}

\end{document}